\documentclass{emulateapj}                                            %

\newcommand{\be}{\begin{equation}}
\newcommand{\ee}{\end{equation}}
\newcommand{\ba}{\begin{eqnarray}}
\newcommand{\ea}{\end{eqnarray}}

\newcommand{\xte}{\mbox{XTE~J1814 }}
\newcommand{\saxj}{\mbox{SAX~J1808.4-3658 }}
\newcommand{\herx}{\mbox{Her~X-1 }}

\slugcomment{Accepted by ApJ: September 26, 2008}
\shorttitle{The Properties of XTE J1814-338}

\begin{document}

\title{Constraints on the Properties of the Neutron Star XTE~J1814-338 from Pulse Shape Models}
\author {Denis A. Leahy\altaffilmark{1}, Sharon M. Morsink\altaffilmark{2},
Yi-Ying Chung\altaffilmark{3}, \& Yi Chou\altaffilmark{3}}

\altaffiltext{1}{Department of Physics and Astronomy, University of Calgary,
Calgary AB, T2N~1N4, Canada; leahy@ucalgary.ca}
\altaffiltext{2}{Department of Physics,
University of Alberta, Edmonton, AB, T6G~2G7, Canada; morsink@phys.ualberta.ca}
\altaffiltext{3}{Graduate Institute of Astronomy, National Central University, Jhongli 32001,
Taiwan; yichou@astro.ncu.edu.tw}

\begin{abstract}

The accretion-powered (non-X-ray burst) pulsations of 
XTE J1814-338 are modeled to determine neutron star parameters and their uncertainties. 
The model is a rotating circular hot spot and includes: (1) an isotropic
blackbody spectral component; (2) an anisotropic Comptonized spectral component;
(3) relativistic time-delays and light-bending; and (4) the oblate shape of the star 
due to rotation. This model is the simplest possible
model that is consistent with the data. The resulting best-fit parameters of the model
favor stiff equations of state, as can be seen from the 3-$\sigma$
allowed regions in the mass-radius diagram. We analyzed  all data combined from a 23 day period 
of the 2003 outburst, and separately analyzed data from 2 days of the outburst. 
The allowed mass-radius regions for both cases only allow equations of state (EOS) 
that are stiffer than EOS APR \citep{APR}, consistent with
the large mass that has been inferred for the pulsar NGC 6440B \citep{Fre08}.
The stiff EOS inferred by this analysis is not compatible with the soft EOS
inferred from a similar analysis of SAX J1808.
\end{abstract}


\section{Introduction}
\label{s:intro}

The accretion-powered millisecond-period X-ray pulsars are promising targets for
constraining the neutron star equation of state (EOS) through the modeling of
emission from hot spots on the pulsar's surface. 
The first pulsar discovered in this class, \saxj \citep{Wij98}, has a spectrum
consistent \citep{Gie02} with emission from a hot spot on the star's surface.
Pulse shape modeling of rapidly rotating neutron stars relies on two relativistic
effects: the gravitational bending of light rays reduces the modulation of 
the pulsed emission and depends on the mass to radius ratio $M/R$; and the
Doppler boosting due to the star's rotation creates an asymmetry in the 
pulse shape and depends on the star's radius $R$. These features, combined with
reasonable models of the emission properties at the neutron star's surface can
be used to constrain the neutron star's mass and radius and hence the EOS
of supra-nuclear density matter. 

XTE J1814-338 (hereafter XTE~J1814) was discovered during outburst in June 2003 \citep{MS03}, and
is an accretion powered millisecond pulsar with spin frequency 314.36 Hz
and orbital period of 4.3 hr \citep{Metal03}. A detailed timing analysis
for \xte was performed by \citet{Pap07} to obtain accurate values for
orbital period, projected semi-major axis, pulse spin frequency and spin down rate.
A similar analysis of the pulse arrival times was carried out by \citet{Wat06} and \citet{Chu07}, 
which both included an analysis of phase lags. Soft lags were found in the 2-10 keV energy
band, similar to those for SAX J1808-3658 and consistent with  an origin in Doppler boosting 
of a Comptonized pulse component
with a much broader emission pattern than the blackbody component.

\citet{Str03} found the same frequency in the X-ray bursts as was found in
the persistent emission, but with a lower second harmonic content. \citet{Wat08}
showed that the X-ray burst oscillations are tightly phase-locked with the
non-burst pulsations.
\citet{Bhat05} modeled the oscillations during X-ray bursts with a hot spot model
for a spherical star and for 2 equations of state. Using a large grid of models they 
found an upper limit on compactness $R_{S}/R<0.48$, with $R_{S}$, the Schwarzschild radius.

There are pulse shape models for a few other X-ray pulsars. The 1.2 s period X-ray pulsar
\herx was modeled by \citet{Leahy04} using a model that includes accretion columns. 
The model for \herx constrains the neutron star EOS to a fairly moderate stiffness \citep{Leahy04}.
\citet{Zav98} and \citet{Bog07} have modeled the X-ray emission from the 5.8 ms period radio pulsar PSR~J0437-4715
using a Hydrogen atmosphere model. In the case of PSR~J0437-4715, 
\citet{Bog07} found that a simple isotropic blackbody model is inconsistent with the data.
In their models, \citet{Bog07} showed that the radius of PSR~J0437-4715 
must be larger than 6.7 km if the mass is $1.4 M_\odot$. Unfortunately
the mass of this pulsar is not well-constrained. \citet{Bog08} have shown that constraints on
radius for a number of other ms radio pulsars are also possible. 

Constraints on \saxj (with a spin period of 2.5 ms) were made by \citet{PG03}
 using data from the 1998 outburst. The modeling
done by \citet{PG03} included blackbody emission from a hot spot that is Compton scattered by
electrons above the hot spot. Their model makes use of a spherical model for the star's surface
and does not include the effects of relative time-delays caused by the different time of flights
for photons emitted from different parts of the star's surface. More recently \citet{CLM05}
and \citet{CMLC07} have shown that time-delays and the star's oblate shape are important factors
that can affect the outcome of pulse-shape modeling for rapidly rotating pulsars such as \saxj.
The 1998 outburst data for \saxj was revisited using a models that included time-delays
and oblateness \citep{LMC08} with the result that the EOS for \saxj is constrained to be very soft. 

In this paper we model the accretion-powered pulsations of \xte using a hot-spot model.
The hot-spot model allows for one or two circular hot spots with a two-component spectrum. The
spectral model includes isotropic blackbody emission and an anisotropic Compton-scattered
component described by a powerlaw. The photons are propagated to the observer using the
oblate Schwarzschild approximation \citep{MLCB07} which allows the photon initial conditions to
be placed on an oblate-shaped initial surface determined by an empirical formula. The Schwarzschild
metric is used to compute the photon bending angles and time delays since it has been 
shown \citep{CMLC07} that the 
corrections induced by the Kerr black hole metric or a numerical metric for a rotating star 
are insignificant 
compared to the corrections induced by the oblate shape. In order to do the pulse-shape 
modeling, we 
construct light curves in two narrow energy bands, 2-3 keV and 7-9 keV. We first analyse 
a composite pulse-shape
constructed from 23 days of data and then consider pulse-shapes constructed from single 
days of data in order
to determine whether variations of the pulse-shape with time are significant. 

The outline of this paper is as follows. In Section \ref{s:method} the
 method used to construct the light curves
and analyse them is outlined. The results of the best-fit models are presented in Section \ref{s:results}.
A discussion of the results is presented in Section \ref{s:discuss}. 

\section{Method}
\label{s:method}

\subsection{Construction of Light Curves}

Pulse shapes for the accretion-powered pulsations are constructed
using the ephemeris given by \citet{Chu07}. Data is limited to 
the first 23 days of the 2003 outburst, June 5 to June 27 2003 (MJD 2452795-2452817), in order to 
avoid the later period of the outburst when the flux and pulse shape 
became more erratic (see for example, \citet{Wat05}).
X-ray bursts were cut out of the data during the interval between 100 s
before and after the start of each burst.

\begin{figure}
\plotone{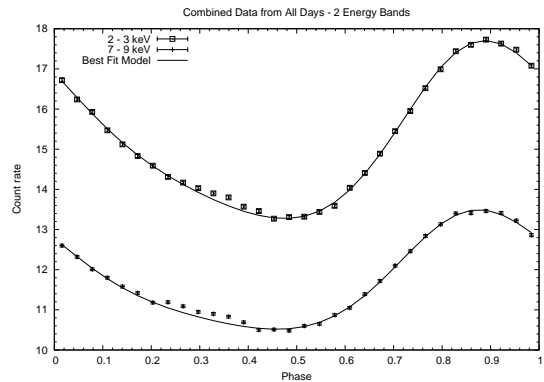}
\caption{Pulse profiles for \xte constructed with data from 
all days between June 5 - 27 (excluding X-ray bursts). Light curves
for two energy bands, 2-3 keV and 7-9 keV are shown.
}
\label{fig:all-days}
\end{figure}

Although the RXTE observations include data in the range of 2 - 50 keV,
we have chosen to concentrate on the lower energy range from 2 - 10 keV for
two reasons. First, the data is noisier at energies above 10 keV. Second,
the Chandra observations by \citet{Kra05} constrain the spectrum in the 
2 - 10 keV range. 
It is also useful to separate the data into narrow energy bands 
in order to separate the different spectral components. We have chosen
two narrow bands, the 2-3 keV band and the 7-9 keV band based on the Chandra
spectrum. The  narrow-band pulse shapes constructed using data
from the full observation period (June 5 - 27) are shown in 
 Figure~\ref{fig:all-days}.

\begin{figure}
\plotone{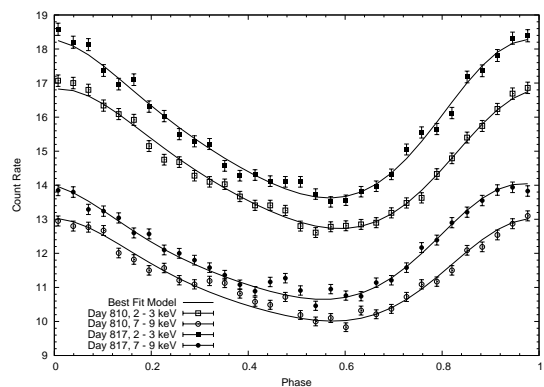}
\caption{One-day pulse profiles for \xte constructed with data from 
days June 20 and June 27 (MJD 2452810 and 2452817). Two energy bands are shown for each day. 
}
\label{fig:2-days}
\end{figure}

We have also investigated the variability of the pulse shape with time.
In order to do this, the data was separated into one-day segments and
separate light curves constructed for each day. It is not computationally
feasible to model all days simultaneously, so we instead focus on two 
separate days. The days were chosen by comparing light curves in the 2 - 10 keV range
for different days using a $\chi^2$ test and selecting two days which differ the most from each other.
This also has the effect of selecting days with intrinsically smaller error bars.
The days resulting from this selection process correspond to June 20 and 27. Light curves for these
two days, in the two narrow energy bands are shown in Figure \ref{fig:2-days}.

\subsection{Analysis of Light Curves}
\label{s:spectrum}

\citet{Kra05} observed \xte with Chandra on June 20, 2003. They modeled the spectrum in the 0.5 - 10 keV
range and found that the
best fit solution corresponds to a combination of a 0.95 keV blackbody and powerlaw emission with a photon 
spectral index of $\Gamma = 1.4$. The ratio of flux from the  blackbody to the powerlaw in their model 
is about 10\%. We use the \citet{Kra05} spectral model and assume that it holds for the other days covered 
by the RXTE data. This assumption is motivated by the fact that the relative normalization of different 
energy bands is approximately constant from day to day, although the overall flux at all wavelengths changes
with time. 

The spectral model of \citet{Kra05} motivates the use of two narrow bands in our pulse shape models. 
A low energy band is necessary in order to capture the blackbody component, so we choose the lowest possible
XTE energy band at 2 - 3 keV. The spectrum in this band is dominated by the powerlaw component, but the
blackbody contribution is still important. We also choose the 7 - 9 keV band as the highest energy band
covered by the Chandra observation. In this high energy band the blackbody flux is negligible. 

Our method for modeling the observed emission is very similar to the method presented in 
\citet{LMC08}. The spectral model has three components: (1) Comptonized flux in the high energy band
(7-9 keV); (2) Comptonized flux in the low energy band (2-3 keV); and (3) blackbody flux in the
low energy band. The observed flux for the $i$th  component, $F_i$, integrated 
over the appropriate observed energy band  is given by 
\be
F_i(E) = I_i \eta^{3+\Gamma_i} (1 - a_i \mu).
\label{eq:flux}
\ee
In equation (\ref{eq:flux}), $I_i$ is a constant amplitude, $\eta$ is the Doppler boost factor,
$\Gamma_i$ is the photon spectral index in the star's rest frame, $\mu$ is the cosine of the
angle between the normal to the star's surface and the initial photon direction, and the
constant $a_i$ describes the anisotropy of the emitted light. For a definition of $\eta$
as well as a more complete description of the modeling method, please see \citet{LMC08}.

In our modeling, the amplitudes $I_1$ and $I_2$ are free parameters  
while the third amplitude 
$I_3$ is defined through the constant $b = \bar{F_3}/\bar{F_2}$, the ratio of the 
phase-averaged blackbody to Comptonized flux in the low-energy band. In the spectral
model by \citet{Kra05}, $b=0.1$, but we include this parameter as a fitting parameter
with 1$\sigma$ limits from their spectral model. 
The photon spectral indices for the Comptonized
components are fixed at $\Gamma_1=\Gamma_2 = 1.4$ as given by their model. In the
narrow range of the low-energy band the blackbody component of $0.95$ keV can be
modeled by a powerlaw with photon spectral index $\Gamma_3=0.85$. The
anisotropy parameters for the Comptonized components $a_1 = a_2 = a$ are assumed to 
be equal, and the parameter $a$ is kept as a free parameter.


In the modelling of the non-accreting ms pulsars, it
was found \citep{Zav98,Bog07,Bog08} that a limb-darkened 
Hydrogen atmosphere spectral model is required by the data. It is 
reasonable to expect that that the blackbody component of the
spectrum should also be limb-darkened. We tested this hypothesis
by multiplying the blackbody flux by a limb-darkening function of the form $e^{-\tau/\mu}$.
We then computed the bestfit neutron star models for two type of models: (1) models with
non-zero optical depth $\tau$ and (2) models with zero optical depth. The bestfit 
models for these two cases are almost identical: the mass and radius of the bestfit model
changes by less than 0.5\% when a nonzero optical depth is added, and the value of $\delta \chi^2 = 0.1$
when the limb-darkening is added. Since the change in $\chi^2$ and the physical parameters are negligible
we conclude that adding an extra parameter to model limb-darkening is not warranted by the data. The 
reason for this is due to the Chandra model which restricts the blackbody contribution in the 2-3 keV
band to only 10\% of the Comptonized contribution, and effectivly sets the blackbody component to zero
in the high energy band. Since the Comptonized flux is dominant and has fan-beaming included, small changes
to the anisotropy of the blackbody component don't affect the final models. For this reason we have
set the anisotropy parameter for the blackbody component to zero ($a_3=0$). This is consistent with the
results found for \saxj \citep{LMC08} which also did not require any limb-darkening.  The final set of free
parameters describing the spectrum are $I_1$, $I_2$, $b$ and $a$.

In order to fit a set of light curves we also need to introduce a set of parameters
describing the star and the emission geometry. These parameters are the mass $M$
and equatorial radius $R$ of the star, the co-latitude of the spot $\theta$,
the inclination angle $i$ as well as a free phase $\phi$. The radius of the 
circular spot (in the star's rest frame) is kept fixed at 1.5 km, as
given by the Chandra spectral model \citep{Kra05}. 

Our models make use of light curves for two different days' data, which requires a 
separate set of parameters for each day. However, on the two different days the 
parameters $M$, $R$ and $i$ do not change. In order to simplify the analysis,
we also assume that the photon spectral indices and the parameters $a$ and $b$
are also fixed. The full set of free parameters are:
$\{I_1, I_2, \theta, \phi\}$ for each day plus $M$, $R$, $i$, $a$ and $b$,
for a total of 13  free parameters. However, for each of our fits, the ratio
$M/R$ is kept fixed, so for any one value of $M/R$, there are only 12 free
parameters.

We use the oblate Schwarzschild approximation \citep{MLCB07} to connect
photons emitted at the star's surface with those detected by the observer.
In previous studies \citep{CLM05,CMLC07} we have shown that, to the 
accuracy required for extracting the parameters of a rapidly rotating 
neutron star, it is sufficient to use the Schwarzschild metric to 
compute the bending of light rays and the relative time delays of 
photons emitted at different locations on the star. The extra time
delays and light bending caused by frame-dragging or higher order
rotational corrections in the metric are negligible. However, the
rotation of the star causes a deformation of the star into an oblate shape,
which changes (relative to a sphere) the directions that photons can 
be emitted into. We have developed a simple approximation \citep{MLCB07}
that allows an empirical fit to the oblate shape of a rotating star to
be embedded in the Schwarzschild metric and make use of it in this 
analysis.

\section{Results}
\label{s:results}

\subsection{Evidence for a Second Spot}
\label{s:bump}

The pulse profiles in Figure \ref{fig:all-days} show a feature
in the phase interval between 0.24 and 0.4. This feature
is seen in all of the other energy bands as well. In order to 
investigate the nature of this feature, we restrict the analysis here to 
just the 7-9 keV light curve. Since the blackbody contribution in this
energy band is negligible, we use a simplified model which only includes
the Comptonized component of the radiation. 

\begin{figure}
\plotone{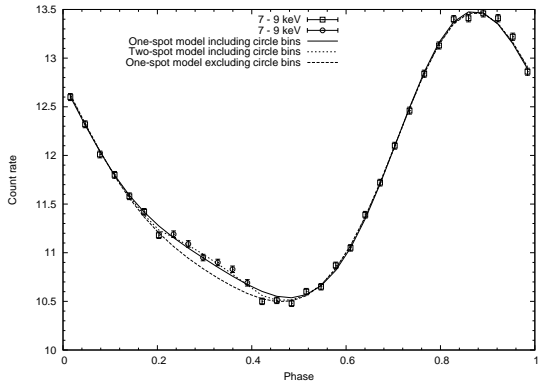}
\caption{Models used to fit the data in the 7-9 keV 
band (data combined from all 23 days). A one-spot model
that uses all data points (both squares and circles) results
in the best fit solid curve with $\chi^2/\hbox{dof} = 50.7/25$.
A two-spot model that uses all data points results in the best fit
dotted curve with $\chi^2/\hbox{dof} = 28.3/23$.
A one-spot model that omits the circle bins results in
the best fit dashed curve with $\chi^2/\hbox{dof} = 19.2/19$.  
}
\label{fig:bump}
\end{figure}

The simplest model for the emission is a single spot. We fitted
the 7-9 keV light curve shown in Figure \ref{fig:bump} with a single
spot model by first fixing $2M/R=0.4$ and varying the parameters $M,i,
\theta, a, I$ and $\phi$. (Similar results are obtained for other
values of $2M/R$.) Since there are 32 data points this corresponds to
25 degrees of freedom. The best fit solution for a single spot model has
$\chi^2 = 50.7$, which is not a very good fit. This best-fit solution is 
shown as a solid curve in Figure \ref{fig:bump}.

We now turn to a two-spot model, where the second spot is allowed to have
an arbitrary location relative to the first spot, but the spectrum
is assumed to be the same. The introduction of a second spot 
introduces three new parameters to the model: an intensity and two
angles. The resulting fit  for $2M/R=0.4$ has $\chi^2 = 28.3$ for 23 degrees of 
freedom, which is a significant improvement. The light curve for this 
model is shown with a dotted curve in Figure \ref{fig:bump}. 
The mass for this model is 2.04 $M_\odot$.
In this model, the
second spot's location is situated so that the second spot is
almost never seen and its light only contributes during the
phase interval between 0.24 and 0.4. Outside of this
interval the light received only originates from the primary
spot. This suggests a simpler one-spot model where bins
in the phase interval 0.24 to 0.4 (marked with
a circle in Figure \ref{fig:bump}) are removed from the data
set. This reduces the degrees of freedom to 19 (32-6 data
points and 5 parameters for a one-spot model). The resulting 
best-fit model for $2M/R=0.4$  has $\chi^2 = 19.2$, which is also
a significant improvement from the one-spot model that uses
all of the data. The mass for this model is $2.08 M_\odot$
and the light curve for this model is shown as a dashed line
in Figure \ref{fig:bump}.

Comparing the two-spot model with the one-spot model with
the circle-bins excluded, we see that the difference in
$\chi^2$ is not significant, and there is little change
between the best-fit values of mass and radius. This 
leads us to the conclusion that there is good evidence
for a second spot (or a feature that mimics a second spot),
but that the amount of data encoded in those bins affected by the
second spot is not sufficient to allow us to model the details of
the second spot with any confidence. Since the
inclusion or exclusion of the second spot does not 
change the best-fit values of the main physical 
parameters of the star ($M, R, i$) it is more
appropriate to choose the simpler one-spot model. 
The qualitative results do not change when we 
look at different energy bands. As a result,
for the remaining modeling reported in this paper
we only use one-spot models where data in the 
0.24 - 0.4 phase range is removed from the analysis.

\subsection{Best-fit Models using Data from All Days}

Our procedure for modeling the two-energy band data shown in 
Figure \ref{fig:all-days} is to assume a one-spot model 
that includes both blackbody and Comptonized emission.
Data in the phase period between 0.24 and 0.4 is
omitted, as described in section \ref{s:bump}. We do a 
number of fits, each with a fixed value of $2M/R$. (Fixing
the ratio of $M/R$ simplifies the fitting procedure since
the light-bending and time delays depend on this ratio.)
Once $2M/R$ is fixed, all other parameters are allowed to vary
and the minimum value of $\chi^2$ is found. In addition to the
parameters described in Section \ref{s:spectrum}, we also added
two parameters corresponding to DC offsets for the two energy
bands. This allows for small errors in the background subtraction.
Once $2M/R$ is fixed, we have a total of 10 free parameters:
$M$, $\theta$, $i$, $a$, $b$, 2 amplitudes, 2 DC offsets
and one overall phase. (The parameter $b$ is restricted to have
a value that is within 1 $\sigma$ of the value found by 
\citet{Kra05}.) Since each energy band has 32 points,
but we exclude 6 of these points we have $64-12-10 = 42$  
degrees of freedom.

For a fixed value of $2M/R$, we find that there are two
local minima. These two minima are shown in Table~
\ref{tab:minima} and we label these two best-fit solutions
as the high and low mass solutions. The lowest value of $\chi^2$
corresponds to the high mass solution and the lower mass solution
has a higher value of $\chi^2$. Although the high mass solution
is a better fit, we exclude this solution on physical grounds.
First of all, it requires a neutron star radius of 28 km, which is 
not allowed by any known equation of state. Secondly, once the
neutron star mass and the inclination angle are known, the 
companion's mass can be calculated (shown in the column labeled 
$M_c$ in Table~\ref{tab:minima}). In the high mass case
the companion's mass is $1.7 M_\odot$. Due to the dim nature
of the companion, \citet{Kra05} have shown that the companion
(if a main sequence star) would have to have a mass that is
no bigger than $0.5 M_\odot$. Clearly this excludes the
high mass solution but allows the lower mass solution.
For these reasons, we exclude the high mass solutions. 

\begin{deluxetable}{lrrrrrrrr}
\tablecaption{Comparison of the Two Minima.\label{tab:minima}}
\tablewidth{0pt}
\tablehead{
\colhead{Model} &
\colhead{$2M/R$}&\colhead{$M$}&\colhead{$R$}&\colhead{$\theta$}&
\colhead{$i$}&\colhead{$a$}&\colhead{$M_c$}&\colhead{$\chi^2/$dof}\\
\colhead{}&\colhead{}&\colhead{$M_\odot$}&\colhead{km}&\colhead{deg.}&
\colhead{deg.}&\colhead{}&\colhead{$M_\odot$}&\colhead{}
}
\startdata
High Mass & 0.3 & 2.86 & 28.7 & 66.8 & 11.8 & 0.81 & 1.70 & 55.9/42 \\ 
Low Mass  & 0.3 & 1.95 & 19.8 & 42.4 & 25.0 & 0.59 & 0.55 & 61.0/42 
\enddata
\end{deluxetable}

In Table \ref{tab:all} the best-fit solution for each value of $2M/R$
is shown. In each case only the lower mass solution is shown. The
best-fit solution shown as a solid curve in Figure \ref{fig:all-days}
corresponds to the $2M/R=0.4$ solution shown in the Table \ref{tab:all}.
Although we call this these solutions ``lower mass'', clearly they
still correspond to high mass neutron stars. In Table \ref{tab:all}, 
only solutions for $2M/R = 0.3$ to $0.6$ are shown. In the case of less
compact stars, such as $2M/R = 0.2$, the only solution corresponded to the
``high mass'' branch of unphysical solutions. We did not test solutions that
were more compact than $2M/R=0.6$ since these solutions would allow spots to 
have multiple images, and our program is unable to handle multiple images.
Technically, the solutions with $2M/R=0.6$ could have spots with multiple
images, but we have checked that the relative values of $\theta$ and $i$
do not lead to this problem for our solutions for the most compact stars.

\begin{deluxetable}{rrrrrrrr}
\tablecaption{Best-fit solutions using data from all days.\label{tab:all}}
\tablewidth{0pt}
\tablehead{
\colhead{$2M/R$}&\colhead{$M$}&\colhead{$R$}&\colhead{$\theta$}&
\colhead{$i$}&\colhead{$a$}&\colhead{$M_c$}&\colhead{$\chi^2/$dof}\\
\colhead{}&\colhead{$M_\odot$}&\colhead{km}&\colhead{deg.}&
\colhead{deg.}&\colhead{}&\colhead{$M_\odot$}&\colhead{}
}
\startdata
0.3 & 1.95 & 19.8 & 42.4 & 25.0 & 0.59 & 0.55 & 61.0/42 \\ 
0.4 & 2.45 & 18.4 & 47.0 & 24.2 & 0.61 & 0.65 & 61.2/42 \\ 
0.5 & 2.38 & 14.2 & 36.7 & 39.4 & 0.59 & 0.39 & 61.9/42 \\ 
0.6 & 2.42 & 11.9 & 40.9 & 42.8 & 0.59 & 0.37 & 62.6/42 \\ 
\enddata
\end{deluxetable}

\begin{figure}
\plotone{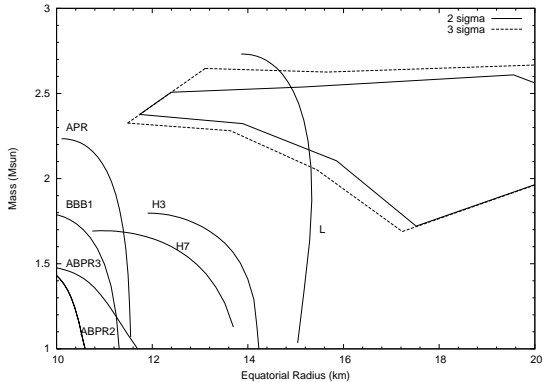}
\caption{Best-fit mass and radius values using the combined data from all days,
separated into two narrow energy bands. Contours shown are for 2- and 3-$\sigma$
confidence levels. Mass-Radius curves for stars spinning at 314 Hz are shown 
as solid curves. The EOS shown are: APR \citep{APR}, BBB1 \citep{BBB},
ABPR2-3 \citep{ABPR}, H3-7 \cite{LNO06} and L (mean-field theory, pure 
neutrons \citep{PPS76}).
}
\label{fig:b1}
\end{figure}

In Figure \ref{fig:b1} we show a number of mass-radius curves
for stars spinning at 314 Hz as well as the 2- and 3-$\sigma$ confidence
regions for the ``low mass'' solutions. (In this figure, the radius $R$
refers to the equatorial radius.) These regions are found by 
fixing the value of $2M/R$, and varying all other parameters and finding
the solutions that have $\chi^2$ larger than the global 
minimum of $\chi^2_{min}=61.0$ by
$\delta \chi^2 = 4$ (2 $\sigma$) or 9 (3 $\sigma$).
The region allowed with 99.7\% confidence (3 $\sigma$) only includes
large stars with high mass. Of the equations of state displayed in
Figure \ref{fig:b1}, the only one lying in the 3 $\sigma$ allowed
region is L, corresponding to pure neutron matter computed in
a mean field approximation \citep{PPS76}. A pure neutron core
is unlikely to be the correct description of supra-nuclear
density matter. However it is possible for an EOS that includes
some softening due to the presence of other species to be allowed
by this data.

\subsection{Best-fit Models Using Data From 2 Days}

It is possible for variability of the data to affect the fit results,
as appears to be the case for SAX J1808 \citep{LMC08}. For this reason
we have rebinned the data into one-day segments in order to see if there
is any significant change in the pulse shape on a day to day basis. 
We performed a $\chi^2$ test to see how closely each day's data matched 
the other days' data. Comparisons of one day with an adjacent day 
gave values of $\chi^2/\hbox{dof}$ ranging from 0.8 to 1.6, indicating 
day-to-day changes are small. The largest change is between
day 810 (June 20) and day 817 (June 27) with $\chi^2/\hbox{dof}$ = 4.8.
Light curves in the two narrow energy bands for these two days are shown
in Figure \ref{fig:2-days}. 

The data corresponding to these two days is binned into 32 bins per period.
Since we continue to remove the 6 bins corresponding to the second ``spot'',
this corresponds to a total of 104 data points.
We fit the data for these two days by assuming that the parameters 
$M, R, i, a$ are the same for both days. The spot's latitude is allowed to
vary, as are the amplitudes of the energy bands and the value of phase. 
Since the DC contributions found in our previous fits are very small, we
do not include terms for DC offsets. In addition, we keep the value of $b$ 
fixed at the \citet{Kra05} value in order to simplify the fits. This corresponds
to 12 parameters, and a total of 92 degrees of freedom. 

\begin{figure}
\plotone{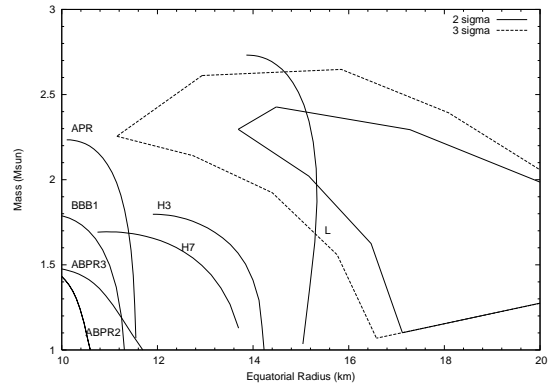}
\caption{Best-fit mass and radius values using data only from day 810 and 817,
separated into two narrow energy bands. Contours shown are for 2- and 3-$\sigma$
confidence levels. EOS labels are the same as in Figure \ref{fig:b1}.
}
\label{fig:a}
\end{figure}

The best-fit solutions for this 2-day joint fit are shown in  Table \ref{tab:2days}.
For each fixed value of $2M/R$ we only found one minimum, unlike the case with 
all days included. The angular locations of the spot on the two days are 
are labeled $\theta_1$ and $\theta_2$.
The change in angular location of the spot between the
two days is less than $2^\circ$ in all cases. The solutions continue to have 
large masses and radii, as in the case of fits using all of the data. In the
case of $2M/R= 0.2$ a low mass (1.2 $M_\odot$) solution is allowed, but it has a very large radius.
The 2- and 3-$\sigma$ confidence regions for the two-day joint fits are shown
in Figure \ref{fig:a}. The 3-$\sigma$ confidence region is somewhat larger than
the same region computed using all of the data, but the two methods have a significant
overlap. The 2-day joint fit also only allows the stiffest EOS L.

\begin{deluxetable}{rrrrrrrrr}
\tablecaption{Best-fit solutions using data only from Days 810 \& 817.\label{tab:2days}}
\tablewidth{0pt}
\tablehead{
\colhead{$2M/R$}&\colhead{$M$}&\colhead{$R$}&\colhead{$\theta_1$}& \colhead{$\theta_2$} &
\colhead{$i$}&\colhead{$a$}&\colhead{$M_c$}&\colhead{$\chi^2/$dof}\\
\colhead{}&\colhead{$M_\odot$}&\colhead{km}&\colhead{deg.}& \colhead{deg.} &
\colhead{deg.}&\colhead{}&\colhead{$M_\odot$}&\colhead{}
}
\startdata
0.2 & 1.18 & 18.4 & 33.6 & 34.7 & 33.8 & 0.55 & 0.29 & 123.9/92 \\ 
0.3 & 1.71 & 17.3 & 34.1 & 35.3 & 32.0 & 0.55 & 0.39 & 124.6/92 \\ 
0.4 & 2.13 & 16.1 & 35.1 & 36.6 & 33.6 & 0.55 & 0.43 & 125.4/92 \\ 
0.5 & 2.41 & 14.4 & 37.1 & 38.7 & 36.3 & 0.55 & 0.43 & 127.5/92 \\ 
0.6 & 2.55 & 12.6 & 39.5 & 41.4 & 39.9 & 0.56 & 0.41 & 132.0/92 \\ 
\enddata
\end{deluxetable}

\subsection{Dependence of Models on Assumed Parameters}

The models in this paper depend on the results of the spectral models of
\citet{Kra05}. We now consider the effect of allowing the parameters in the
spectral model to vary within the error bars. As mentioned in Section~\ref{s:spectrum} we 
already allow the ratio of the blackbody to powerlaw components ($b$) to vary within the 
1$\sigma$ limits given by the \citet{Kra05} spectral model. 
Another spectral parameter that could 
affect the fits is the photon spectral index $\Gamma$ for the powerlaw 
component of the spectrum. \citet{Kra05} found a value of $\Gamma = 1.41 \pm 0.06$
and in all of our models presented in the previous section we kept the photon spectral
index fixed at a value of $\Gamma = 1.40$. We would expect that a larger value of $\Gamma$
 would allow for smaller stars. This is because the flux in Equation (\ref{eq:flux}) is 
proportional to the Doppler boost factor $\eta$ raised to the power $\Gamma + 3$. 
The Doppler boost factor is mainly responsible for introducing higher harmonics into the
signal, so a larger value of $\Gamma$ creates a more asymmetric pulse shape. In order to 
compensate, the best-fit solution will require a smaller value of stellar radius $R$ 
in order to decrease the value of $\eta$. In order to test the dependence of the best-fit
values of the parameters on $\Gamma$, we chose a value of $\Gamma=1.50$ which is somewhat
larger than the range allowed by \citet{Kra05} and fit the data using the same method
described earlier in this paper. The results of the best-fit parameters for the two
values of $\Gamma$ for the case of $2M/R=0.4$ are shown in Table~\ref{tab:test}. 
As expected, increasing $\Gamma$ allowed for a smaller star, but the decrease is
only by 3\%. Similar results occur for other values of $M/R$. Clearly the 
dependence on the photon spectral index is not sensitive enough to affect the 
resulting large size of the best-fit stars.

In our models we keep the 
spot size (as measured on the star's surface) fixed at a diameter of 3 km. In previous
modeling \citep{LMC08} of \saxj we found that the final values of the best-fit
parameters were not sensitive to changes in the spot size, assuming that the spot is 
small compared to the size of the star. For this reason we have kept the spot size
fixed at 3 km for all models in this analysis. We have also assumed the hot spot
is circular, although recent MHD models \citep{KR05} more complicated spot shapes.
In this paper we have only attempted to use the simplest possible model that is
still consistent with the data. 
Adding extra parameters to our models in order to describe more complicated spot
shapes is not yet warranted by the quality of the data.

For all models computed so far, we have made use of an empirical formula for the 
oblate shape of the star, and have included the relative time-delays for photons
emitted from different parts of the star. In Table \ref{tab:test} the 
effects of oblateness and time-delays on the fits are shown. For the model labeled ``sphere'',
a spherical initial surface was assumed, but relative time-delays were included
in the computation. The resulting best-fit solution is about 10\% larger than
the corresponding oblate model (labeled ``oblate'' and $\Gamma=1.4$ in Table 
\ref{tab:test}). This shrinkage of the star's radius when oblateness is
included has been observed in the modeling of \saxj \citep{LMC08}. 
For the model labeled ``no td'' time-delays were omitted from the calculation
and a spherical surface was used.
Comparison of the two spherical models
in Table \ref{tab:test} shows that the model that includes time-delays
is about 3\% smaller than the model that omits time-delays. 

\subsection{Comparison with X-ray Burst Data}

In our analysis of \xte we have only included data from the accretion-powered
pulsations and have omitted any data corresponding to an X-ray burst. \citet{Bhat05}
analyzed the light curves constructed from the X-ray bursts for this neutron star.
In their analysis they assumed a spherical surface for the star and traced the 
paths of the X-rays using the Kerr metric. They also made use of a limb-darkened
blackbody emission (2 keV) spectral model appropriate for X-ray bursts. Due to 
the method that they adopted, it was necessary for them to assume one of two
different equations of state. The stiffer EOS used by \citet{Bhat05} is
the same as the EOS that we label ``APR'' and is the A18+$\delta v$+UIX 
computed by \citet{APR}. The analysis of the X-ray bursts by \citet{Bhat05} allows the APR
EOS, while our analysis of the accretion-powered pulsations only allows stiffer EOS. 
From their analysis it is difficult to determine whether or not their analysis of
the X-ray burst data is consistent with a very stiff EOS, as indicated by our analysis.

\citet{Wat05} provide a detailed analysis of many aspects of both the
X-ray bursts and the non-burst emission. One of the quantities that they
measured was the fractional amplitude of the pulsations at the fundamental
frequency and the first harmonic for both the burst and non-burst emission.
They found that the non-burst 
emission (modeled in this paper) has a larger harmonic content than the
burst emission studied by \citet{Bhat05}. Since Doppler boosting is partially
responsible for the harmonic content, it is perhaps not surprising that our
analysis of the non-burst pulsations implies a larger Doppler factor for the
star, which in turn implies a larger radius than a similar analysis for
the X-ray burst oscillations.

\begin{deluxetable}{lrrrrrrrrrr}
\tablecaption{Dependence of Models on Parameters. Joint fits for two energy bands
for two separate days (810 and 817).\label{tab:test}}
\tablewidth{0pt}
\tablehead{
\colhead{Model}& \colhead{$\Gamma$} &
\colhead{$\frac{2M}{R}$}&\colhead{$M$}&\colhead{$R$}&\colhead{$\theta_1$}& \colhead{$\theta_2$} &
\colhead{$i$}&\colhead{$a$}&\colhead{$M_c$}&\colhead{$\chi^2/$dof}\\
\colhead{}& \colhead{} &
\colhead{}&\colhead{$M_\odot$}&\colhead{km}&\colhead{deg.}& \colhead{deg.} &
\colhead{deg.}&\colhead{}&\colhead{$M_\odot$}&\colhead{}
}
\startdata
oblate & 1.4  &0.4 & 2.13 & 16.1 & 35.1 & 36.6 & 33.6 & 0.55 & 0.43 & 125.4/92 \\ 
oblate & 1.5  &0.4 & 2.07 & 15.6 & 34.5 & 35.9 & 34.3 & 0.55 & 0.41 & 126.6/92 \\ 
sphere & 1.4  &0.4 & 2.38 & 17.6 & 32.8 & 34.3 & 31.6 & 0.54 & 0.49 & 124.7/92 \\ 
 no td & 1.4 & 0.4 & 2.45 & 18.5 & 49.0 & 50.9 & 21.1 & 0.60 & 0.76 & 126.8/92 
\enddata
\end{deluxetable}

\section{Discussion}
\label{s:discuss}

We have analyzed the accretion-powered (non-X-ray burst) pulsations of 
\xte using a hot spot model. Our modeling includes (1) an isotropic
blackbody spectral component; (2) an anisotropic Comptonized component;
(3) relativistic time-delays; (4) the oblate shape of the star due to
rotation. The model presented in this paper is the simplest possible
model that is consistent with the data. The resulting best-fit models
favor stiff equations of state, as can be seen from the 3-$\sigma$
allowed regions in Figures \ref{fig:b1} and \ref{fig:a}. In Figure
\ref{fig:b1} all data from a 23 day period of the 2003 outburst were
included, while for Figure \ref{fig:a} data from only 2 days were 
included. The allowed regions for the two data sets differ slightly,
but both only allow equations of state that are stiffer than 
EOS APR \citep{APR}. 

It is interesting that a large mass has been inferred for the pulsar
NGC 6440B \citep{Fre08} through measurements of periastron precession.
Assuming that the observed periastron precession is purely from
relativistic effects, the pulsar's mass is $M = 2.74 \pm 0.21 M_\odot$
(1-$\sigma$ error bars) \citep{Fre08}. If the mass really is this high,
it would be consistent with the stiff equations of state allowed by
our analysis of \xte. However, it is still possible that the
large periastron precession observed for NGC 6440B 
could be caused by a very rapidly rotating companion \cite{Fre08},
in which case the pulsar's mass would be smaller and compatible with
more moderate equations of state. A high mass for \saxj is also
inferred by observations during its quiescent state \citep{Hei07}. 
Modelling of the neutron star X7 in 47 Tuc \citep{Hei06} allows
for a high mass neutron star, although for X7 a  low mass neutron star 
is also allowed. 

Similar hot spot models of \saxj imply a soft equation of state
and a column model for Her X-1 also implies a soft EOS (see \cite{LMC08}
and \cite{Leahy04} for details). 
The best-fit pulse-shape models found for \xte have mass and radius incompatible
with the 3-$\sigma$ allowed regions of mass and radius 
found for \saxj or Her X-1. Some possible
interpretations could be (1) time-variations in the pulse profile
of \saxj led to an underestimate of the star's radius; (2) the equation of state
for dense matter has a two-phase nature allowing both large and
small compact stars;  or (3) the simple hot-spot model doesn't 
describe one or more of these pulsars.

The first reason is a factor for \saxj.
It was demonstrated in \cite{LMC08} that the analysis of the pulse profile averaged
over a long observation gave significantly different results than the pulse
profile obtained from a much shorter observation. 
Supporting this conclusion is a recent analysis 
by \cite{Har08} of data from the
1998, 2002 and 2005 outbursts of \saxj showing a great deal of
variation in the pulse shape over time (see Figure 3 of \citet{Har08}). 
The pulse-shape analysis by \citet{LMC08} only made use of the data
from the 1998 outburst. The 1998 data is very sinusoidal in nature
and has very little harmonic content. The results of \citet{Har08}
show that the later outbursts have a stronger harmonic content.
Since a larger radius star can produce a stronger harmonic content,
it is possible that the addition of data from 2002 and 2005 will
alter the conclusions of \citet{LMC08} about SAX J1808.
But the effect of pulse shape variability is not
a factor for Her X-1 where the pulse shape has high stability. 

The second reason above, i.e. a bimodal equation of state, is a possible, but 
speculative, solution to the greatly different allowed regions for M and R.  
In this scenario, there is still only one baryonic equation of state and one quark
matter EOS, but above a certain critical density, $\rho_{crit}$, the whole star 
makes a transition from baryonic  matter to quark matter. 
Then for stars with central density $\rho_c$ having $\rho_c<\rho_{crit}$, 
the M vs. R relation follows a stiff baryonic EOS, somewhat like EOS L, whereas
for $\rho_c>\rho_{crit}$ the star has converted to a quark star and lies on a
quark matter EOS curve in the mass-radius diagram. In the case of EOS L,
the $3 \sigma$ region with $M<2.7 M_\odot$ would require a value of 
$\rho_{crit} \sim 10^{15} g/cm^3$.
Since quark matter EOS curves have a lower maximum mass than baryonic EOS curves, 
any baryonic star that makes
the transition and has mass above the quark star maximum mass must lose mass to
end up as a stable quark star. The mass loss depends on the physics of the transition
process and is likely vary from star to star. In this scenario, we interpret 
\xte to lie on the baryonic branch of the mass vs. radius diagram and
\saxj and Her X-1 to lie on the quark matter branch. 

The third listed reason for the discrepancy, that the emission region models are 
too simple to represent the actual emission regions on the stars, is a definite 
possibility. For instance, if the emission is coming from the magnetosphere, then
our models are incorrect. Alternatively, the emission may arise from surface spots, but
the region's shape might be more complicated than a circle.
This can only be tested by constructing more complex models and applying
them to the observed pulse shapes. However with a more complex model with more parameters
describing the model, better pulse shape data is required to constrain the model parameters.
Future work is planned to explore more complex emission models, and to test whether
these resolve the apparent discrepancy in mass and radius for different pulsars.

\acknowledgments
This research was supported by grants from NSERC to D. Leahy and S. Morsink.
Y. Chung and Y. Chou acknowledge partial support from Taiwan National
Science Council grant NSC 95-2112-M-008-026-MY2.
We also thank the Theoretical Physics
Institute at the University of Alberta for supporting D. Leahy's visits
to the University of Alberta.

\end{document}